\documentclass[amsmath,amssymb,twocolumn,amsfonts]{revtex4}
\usepackage{graphicx}
\usepackage{epsfig}
\usepackage{bm}
\usepackage{bbm,mathbbol}
\usepackage{braket,bigints}
\usepackage{stmaryrd,mathtools}
\def \beq {\begin{equation}}
\def \eeq {\end{equation}}
\def \tr {\rm Tr}

\begin{document}
\title{Spin-exchange collisions in hot vapors create and sustain bipartite entanglement}
\author{K. Mouloudakis and I. K. Kominis}
\affiliation{Department of Physics and Institute of Theoretical and Computational Physics, University of Crete, 70013 Heraklion, Greece}

\begin{abstract}
Spin-exchange collisions in alkali or alkali/noble gas vapors are at the basis of quantum sensing, nucleon structure studies, tests of fundamental symmetries, and medical imaging. We here show that spin-exchange collisions in hot alkali vapors naturally produce strong bipartite entanglement, which we explicitly quantify using the tools of quantum information science. This entanglement is shown to have a lifetime at least as long as the spin-exchange relaxation time, and to directly affect measurable spin noise observables.
 This is a formal theoretical demonstration that a hot and dense atomic vapor can support longlived bipartite and possibly higher-order entanglement.
\end{abstract}
\maketitle 
Atomic spin-exchange collisions are at the basis of far-ranging explorations, from nuclear physics \cite{Deur} and astrophysics \cite{Zygelman_2005,Loeb_2012} to quantum sensing \cite{Optical_Magnetometry} and medical imaging \cite{Goodson,Saam}. Spin-exchange collisions in hot alkali vapors underlie the dynamics of optical pumping and spin relaxation, both being central to the production and probing of non-equilibrium magnetic substate populations by light \cite{Happer_RMP_1972,Happer_Primer,Walker_Happer_RMP_1997}. More recently, the intricate properties of spin-exchange collisions \cite{Happer_Tam_PRA_1977,Happer_Book} have spurred the development of ultrasensitive atomic magnetometers \cite{rom1,rom2,rom3}, advancing precision tests of fundamental physics \cite{rom4,rom5,rom6} and biomagnetic imaging \cite{meg1,meg2,meg3,meg4, Weis}. Since relaxation is intimately connected to fluctuations, spin-exchange collisions also generate spin noise, spontaneous fluctuations of the collective spin addressed by spin noise spectroscopy \cite{sn1,sn2,sn3,sn4,sn5,sn6,sn7,MoulPRR}.

The existing theoretical treatment \cite{Grosettete,Happer_Primer,Happer_Book} of alkali atom spin-exchange collisions treats the two atoms emerging from a binary collision as uncorrelated, and can thus account for single-atom observables in an atomic vapor understood as consisting of uncorrelated atoms. This approach has been quite successful, because so far experiments probed mostly single-atom observables. However, recent years have witnessed the exploration of quantum-correlated states in such vapors. In particular, measurement-induced multi-atom entanglement in hot alkali vapors has just recently been experimentally observed \cite{Mitchell}. Moreover, quantum correlations in alkali or alkali/nobel gas vapors have been theoretically discussed in \cite{Kominis2008,Firstenberg1,Firstenberg2}.

Here we address the quantum foundations of alkali-alkali spin-exchange collisions and show that they produce strong bipartite (atom-atom) entanglement. We formally quantify this entanglement with the tools of quantum information science, and show it persists for at least another ten spin-exchange collision times. Hence we provide a formal demonstration, using the full alkali atom density matrix and the full spin-exchange interaction, that random spin-exchange collisions in a hot and dense vapor fundamentally allow for long-lived two-body entanglement. It is conceivable that these results hold true also for multi-body entanglement produced by a sequence of binary spin-exhange collisions. Here we terminate this sequence at the third collision partner, limiting this discussion to bipartite entanglement. We finally show how this entanglement can be revealed through binary spin correlations affecting measurable spin noise variances.  

Spin-exchange (SE) collisions between two atoms, A and B, result from the different potential curves \cite{Kartoshkin} of the singlet and triplet total spin of the colliding partners. If $\mathbf{s}_A$ and $\mathbf{s}_B$ are the electron spins of the colliding atoms, the singlet and triplet projectors are $P_S={1\over 4}-\mathbf{s}_A\cdot\mathbf{s}_B$ and $P_T={3\over 4}+\mathbf{s}_A\cdot\mathbf{s}_B$, respectively \cite{notation}. Introducing the exchange operator $P_e=P_T-P_S$, the SE interaction potential can be written as \cite{Happer_Book} $V_{\rm se}=V_0+V_1P_e$. Only the latter term drives the spin state evolution, expressed by the unitary operator $U_\phi^{\rm AB}=e^{-i\int dt V_1P_e}$, where $\phi=\int dt V_1$ is the SE phase \cite{note0}. Noting that $P_e^2=\mathbb{1}$, we find $U_\phi^{\rm AB}=\cos\phi\mathbb{1}-i\sin\phi P_e\label{U}$. 
Now let two uncorrelated atoms A and B enter a collision in the combined state $\rho_0=\rho_a\otimes\rho_b$. The post-collision density matrix is $\rho=U_\phi^{\rm AB}\rho_0 (U_\phi^{\rm AB})^\dagger=\cos^2\phi\rho_0+\sin^2\phi P_e\rho_0P_e-(i/2)\sin2\phi[P_e,\rho_0]$.

The next step in the standard derivation of SE relaxation \cite{Grosettete,Happer_Primer,Happer_Book} is to trace out atom B (A) in order to find the post-collision state of atom A (B), writing the combined post-collision state as $\rho_a'\otimes\rho_b'$, where $\rho_a'=\tr_B\{\rho\}$ and $\rho_b'=\tr_A\{\rho\}$. With the post-collision state written as a tensor product of uncorrelated states, this approach is well suited for treating single-atom observables in a vapor described as consisting of uncorrelated atoms. Hence there is no need to keep track of the tensor-product notation and one is left with a single-atom density matrix evolution (for single-species vapors one sets $a=b$ and omits the atom indeces altogether).

We will now extend this treatment and unravel the bipartite entanglement in the post-collision two-atom state $\rho$ resulting from the action of $U_\phi^{\rm AB}$ on the initial state $\rho_0$. To facilitate this analysis, we write $\rho$ as \cite{note1} 
\begin{widetext}
\begin{align}
\rho&=\Big(\cos^2\phi+{1\over 4}\sin^2\phi\Big)\rho_a\otimes\rho_b+{1\over 4}\sin^2\phi\sum_j\sigma_j\rho_a\sigma_j\otimes\sigma_j\rho_b\sigma_j\nonumber\\
&+{{\sin^2\phi}\over 4}\Big[\sum_{i\neq j}\sigma_i\rho_a\sigma_j\otimes\sigma_i\rho_b\sigma_j+\sum_j\big(\rho_a \sigma_j\otimes\rho_b \sigma_j+\sigma_j\rho_a\otimes \sigma_j\rho_b\big)\Big]
-{{i\sin2\phi}\over 4}\sum_j\big(\sigma_j \rho_a \otimes \sigma_j \rho_b - \rho_a \sigma_j \otimes \rho_b \sigma_j\big)\label{rho}
\end{align}
\end{widetext}
We will first find a formal upper bound to the entanglement of $\rho$, which bound is independent of the SE phase $\phi$ and the particular colliding states $\rho_a$ and $\rho_b$. This will serve both as an indicative measure of entanglement and as a consistency check for the numerical calculations following suit and demonstrating that several colliding states of practical interest lead to significant entanglement, in cases saturating the upper bound. Incidentally, a general lower bound other than the trivial one (zero) cannot be given, since the entanglement of $\rho$ depends on $\phi$, and for $\phi=0$ or $\phi=\pi$ it is $\rho=\rho_0$, in which case $\rho$ has zero entanglement.

We first note that the first line in Eq. \eqref{rho} is a separable density matrix, i.e. it is written as $\sum_i p_i\rho_a^{(i)}\otimes\rho_b^{(i)}$, with $\sum_i p_i=1$. The form $\rho_a^{(i)}\otimes\rho_b^{(i)}$ is obvious in the first term of the first line in Eq. \eqref{rho}. Regarding the second term, each term in the sum over $j$ multiplied by ${1\over 4}\sin^2\phi$ is a physical tensor-product density matrix, because it is the result of acting on $\rho_a\otimes\rho_b$ with a completely positive map consisting of $\sigma_j$, since $\sigma_j$ is hermitian and $\sigma_j^\dagger\sigma_j=\sigma_j^2=\mathbb{1}$. For example, the term $j=x$ results from acting on $\rho_a\otimes\rho_b$ with $M_x=\sigma_x\otimes\sigma_x$, i.e. from the operation $M_x\rho_a\otimes\rho_b M_x^\dagger$. Since all such $M_j$ operators are local, the resulting density matrices in each of the three such terms ($j=x,y,z$) are again of the separable form. Now, the second line in Eq. \eqref{rho} is hermitian and traceless. If this second line was absent, the density matrix $\rho$ would be separable. But as is, it generally exhibits bipartite entanglement. 

Negativity is an entanglement measure \cite{Plenio} for bipartite systems, defined by ${\cal N}(\rho)=(||\rho^{T_B}||-1)/2$, where $\rho^{T_B}$ is the partial transpose (PT) of $\rho$. The partial transpose of a bipartite density matrix $\rho=\sum_{ijkl}\rho_{ij,kl}\ket{i}\bra{j}\otimes\ket{k}\bra{l}$ is $\rho^{T_B}=\sum_{ijkl}\rho_{ij,kl}\ket{i}\bra{j}\otimes\ket{l}\bra{k}$, i.e. the operator in the right position of the tensor product (party B) is transposed. The trace-norm $||\rho^{T_B}||\equiv\tr\{\sqrt{(\rho^{T_B})^{\dagger}\rho^{T_B}}\}$ is a special case ($p=1$) of the so-called Shatten-p norm. Since the PT of $\rho$ in Eq. \eqref{rho} is hermitian,  $||\rho^{T_B}||$ equals the sum of the absolute values of the eigenvalues of $\rho^{T_B}$.

To analytically calculate $||\rho^{T_B}||$ in all generality presents an insurmountable difficulty. However, we can calculate an upper bound to $||\rho^{T_B}||$, using the triangle  inequality $||A_1+A_2+...A_n||\leq||A_1||+||A_2||+...||A_n||$, and the fact that for any constant c it is $||cA||=|c|||A||$. First note that $\sigma_x^T=\sigma_x$, $\sigma_y^T=-\sigma_y$ and $\sigma_z^T=\sigma_z$. It follows that the 
effect of the PT operation on the first line of Eq. \eqref{rho} is just to change $\rho_b$ into $\rho_b^*$. However, $\rho_b^*$ is also a physical density matrix having opposite phases compared to $\rho_b$. Hence after the PT, the first line is still a physical density matrix of unit trace, and hence its trace norm is 1.

For the terms in the second line of Eq. \eqref{rho} we will use the identity $||A_1\otimes...\otimes A_n||=||A_1||...||A_n||$, and the fact that the trace norm is unitarily invariant, i.e. $||A||=||UAV||$ for unitary $U$ and $V$. We appropriately choose $U$ and $V$ to be some Pauli operator $\sigma_j$, such that we rid all terms in the PT version of the second line of Eq. \eqref{rho} from the $\sigma_j$ operators, also using the fact that $\sigma_j^2=\mathbb{1}$. For example, take the term $A=\sigma_x\rho_a\sigma_y\otimes\sigma_x\rho_b\sigma_y$. Its partial transpose is $A^{T_B}=-\sigma_x\rho_a\sigma_y\otimes\sigma_y\rho_b^*\sigma_x$. The trace norm of $A^{T_B}$ is $||A^{T_B}||=||\sigma_x\rho_a\sigma_y||~||\sigma_y\rho_b^*\sigma_x||$. For the first term in this product we take $U=\sigma_x$ and $V=\sigma_y$, while for the second we choose $U=\sigma_y$ and $V=\sigma_x$. Hence $||A^{T_B}||=||\rho_a||~||\rho_b^*||=1$. We thus reduce all terms to expressions having unit trace norm. There are 12 such terms in the expression multiplied by $(1/4)\sin^2\phi$, and 6 such terms in the one multiplied by $-(i/4)\sin2\phi$. Thus the negativity of $\rho$ given by Eq. \eqref{rho} is bounded by
\beq
{\cal N}(\rho)\leq {3\over 2}\sin^2\phi+{3\over 4}|\sin2\phi|\label{bound}
\eeq
We will next show numerically (apart from two specific examples solved analytically) that states of experimental relevance lead to significant negativities, in cases saturating the bound of Eq. \eqref{bound}. We have performed an exact simulation for a $^{87}$Rb vapor (nuclear spin $I=3/2$, 8-dimensional Hilbert space, 64-dimensional tensor product space). We use random pre-collision states $\rho_a\otimes\rho_b$ and a random SE phase $\phi$. Writing down the most general random, hermitian and positive semidefinite 8-dimensional matrix is not trivial \cite{Petruccione}. Therefore we use random coherent superpositions of the $\ket{FM}$ basis states to create the most general random pure states $\ket{\psi_a}$ and $\ket{\psi_b}$ \cite{note}. 

\begin{figure}[t!]
\begin{center}
\includegraphics[width=7.7 cm]{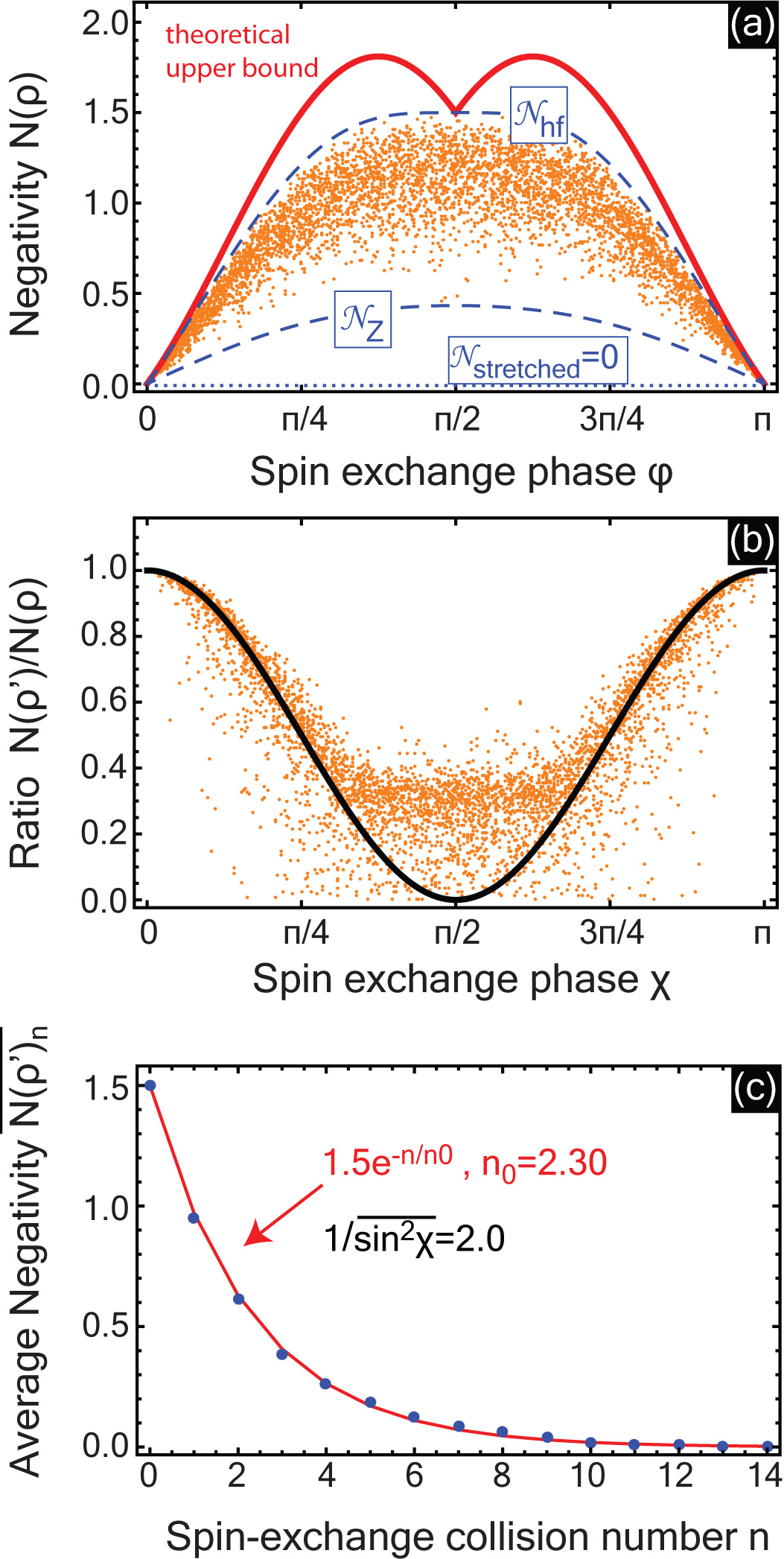}
\caption{(a) Negativity of $\rho$ in Eq. \eqref{rho}. Red solid line is the upper bound of Eq. \eqref{bound}. The 5000 orange dots correspond to random pure initial states $\ket{\psi_a}$ and $\ket{\psi_b}$ for $^{87}$Rb and random SE phase $\phi$. Upper (lower) dashed blue lines are the analytic negativities ${\cal N}_{\rm hf}$ (${\cal N}_{\rm Z}$) for collisions of $\ket{20}$ with $\ket{20}$ ($\ket{22}$ with $\ket{21}$). Dotted curve at zero corresponds to collisions between stretched states, $\ket{22}$ with $\ket{22}$ and $\ket{2-2}$ with $\ket{2-2}$. (b) Ratio of ${\cal N}(\rho')$ with ${\cal N}(\rho)$ for the same states corresponding to the orange dots in (a), where $\rho'$ results from a B-C collision with random phase $\chi$ and random C-atom state $\ket{\psi_c}$, after tracing out atom C. Solid line is $\cos^2\chi$. (c) Evolution of A-B negativity (blue dots), starting out from a highly entangled A-B state with negativity 1.5, then A or B colliding with C atoms randomly chosen among $\ket{2m}$ with random phase $\chi$. Red solid line is an exponential decay with constant $n_0=2.3$, close to $1/\overline{\sin^2\phi}\approx 2.0$ resulting from the Cauchy distribution of $\chi$ having center 0.0 and scale 10.0.}
\label{neg}
\end{center}
\end{figure}

The result is shown in Fig. \ref{neg}a. The largest negativity is produced for collisions of $\ket{20}$ with $\ket{20}$ and $\ket{10}$ with $\ket{10}$, both of which saturate the bound for $\phi=\pi/2$. For those cases, which are relevant to frequency standards \cite{Vanier}, we can find the exact result ${\cal N}_{\rm hf}={3\over 4}(\sin^2\phi+|\sin\phi|\sqrt{1+3\cos^2\phi})$ (dashed upper blue line in Fig. \ref{neg}a). 

On the other hand, collisions between the stretched states $\ket{22}$ with $\ket{22}$, and $\ket{2-2}$ with $\ket{2-2}$ produce zero negativity, ${\cal N}_{\rm streched}=0$. This is because strecthed states are invariant under SE, so an initially uncorrelated product state of stretched states will remain invariant and uncorrelated.

Relevant to a highly spin-polarized vapor are mostly collisions between $\ket{22}$ and $\ket{21}$ states. In fact, such collisions produce the ubiquitous Zeeman frequency shift. Indeed, a {\it perfectly} spin-polarized vapor in the stretched state $\ket{22}$ is invariant under SE, hence there is neither any entanglement nor any Zeeman shift produced. A frequency shift is observable if the vapor is not perfectly polarized. In such a case, assuming the population of $\ket{22}$ is significantly larger than the population of $\ket{21}$, it is collisions between $\ket{22}$ and $\ket{21}$ states that dominate the shift, since collisions of $\ket{21}$ with $\ket{21}$ are less frequent. The negativity for a collision of $\ket{22}$ with $\ket{21}$ (lower dashed blue line in Fig. \ref{neg}a) can also be found analytically, it is ${\cal N}_{\rm Z}={1\over 4}|\sin\phi|\sqrt{3+\cos^2\phi}$. 

Based on Fig. \ref{neg}a and the exact results ${\cal N}_{\rm hf}$ and ${\cal N}_{\rm Z}$, it appears that the entanglement produced by strong SE collisions, the phase of which is such that the collisional average $\overline{\sin^2\phi}\approx 1$ \cite{Happer_Book}, is rather significant. We next turn to explicitly quantify the lifetime of this entanglement. To this end, we will bring into the picture a third atom C, and consider the uncorrelated initial state $\rho_a\otimes\rho_b\otimes\rho_c$. We let atoms A and B collide with  phase $\phi$ as before, and then have atom B collide with atom C with phase $\chi$. We  then trace out atom C, and find the negativity of the resulting A-B state 
\beq
\rho'=\tr_{C}\{U_{\chi}^{BC}U_{\phi}^{AB}\rho_a\otimes\rho_b\otimes\rho_c (U_{\phi}^{AB})^\dagger(U_{\chi}^{BC})^\dagger\}.\label{rho'}
\eeq
From the resulting expression we can ignore terms proportional to either $\sin 2\phi$ or $\sin 2\chi$, the collisional averages of which express the collisional frequency shift and thus are very small \cite{Happer_Book} , and thus we get
\begin{align}
\rho'&\approx \cos^2\chi \rho\nonumber\\
&+\cos^2\phi\sin^2\chi\rho_a\otimes\tr_{C}\{P_e^{\rm BC}\rho_b\otimes\rho_c P_e^{\rm BC}\}\nonumber\\
&+\sin^2\phi\sin^2\chi\tr_{C}\{P_e^{\rm BC}P_e^{\rm AB}\rho_a\otimes\rho_b\otimes\rho_c P_e^{\rm AB}P_e^{\rm BC}\}\label{rho'2}
\end{align}
In Fig. \ref{neg}b we plot the exact ratio ${\cal N}(\rho')/{\cal N}(\rho)$ as a function of $\chi$. It is seen that
${\cal N}(\rho')\approx \cos^2\chi{\cal N}(\rho)$, i.e. the negativity of $\rho'$ is approximately given by considering just the first term in Eq. \eqref{rho'2}. Since a binary SE collision happens every time interval $T$, related to $T_{\rm se}$ by \cite{Happer_Primer} $\overline{\sin^2\chi}/T=1/T_{\rm se}$, and since just one B-C collision reduces the A-B negativity by $\cos^2\chi$, we can write 
\beq
{{d{\cal N}(\rho)}\over {dt}}\approx{{{\cal N}(\rho')-{\cal N}(\rho)}\over T}\approx -{{{\cal N}(\rho)}\over T_{\rm se}}
\eeq
Thus the negativity  ${\cal N}(\rho)$ is predicted to decay exponentially with time constant $T_{\rm se}$. Indeed, this is explicitly shown in the example of Fig. \ref{neg}c. To produce this plot we let both atoms A and B initially collide in the state $\ket{20}$ with phase $\pi/2$, thus producing a highly entangled state with negativity 3/2. We then let either A or B collide with an atom C randomly chosen among the set of states $\ket{2m}$, and with phase sampled from a Cauchy distribution having zero mean and scale 10.0, producing an average $\sin^2\chi\approx 0.5$. We consider in total 15 collisions labelled by $n=0,1,...,14$ ($n=0$ is the initial A-B collision and the rest are A-C or B-C collisions). After each A-C or B-C collision atom C is traced out. We repeat this process for 100 times and plot the resulting average $\overline{{\cal N}(\rho')}_n$ as a function of SE collision number $n$. We find a decay "time" (in terms of number of SE collisions) very close to $1/\overline{\sin^2\chi}$ as determined from the Cauchy distribution of the $\chi$ values. In particular, it is observed that significant negativity ($\geq 0.1$) survives for about $7T_{\rm se}$.  
\begin{figure}[t]
\begin{center}
\includegraphics[width=8.5 cm]{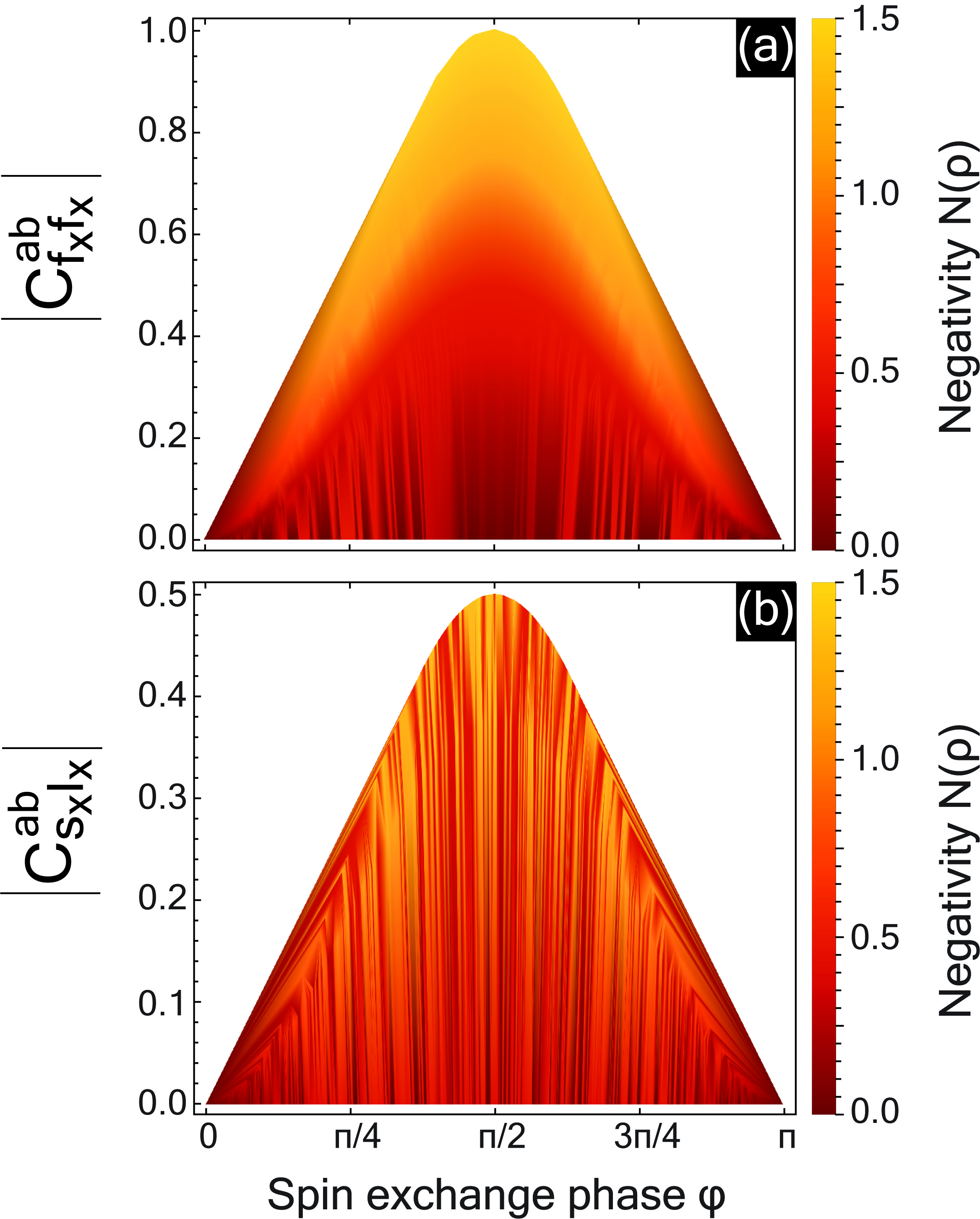}
\caption{Binary spin correlations (a) $C_{f_xf_x}^{ab}$ and (b) $C_{s_xI_x}^{ab}$ versus the negativity of the two-atom state $\rho$ of Eq. \eqref{rho} produced by an SE collision with phase $\phi$ between $\ket{2m}$ and $\ket{2m'}$, with $m,m'=-2,-1,...,2$. As expected, strong binary spin correlations are seen to be connected with large negativities (correlation coefficient 0.89 in (a) and 0.72 in (b)).}
\label{corr}
\end{center}
\end{figure}

It should be stressed that in all of the above considerations we have not specified the precise physical process realizing the tracing out of atom C. But our starting point was the existing single-atom derivation tracing out atom B (and A) when obtaining the single-atom density matrix under the assumption of {\it instant} A-B decorrelation. We found the natural timescale of this decorrelation by going into a deeper layer of the many-body spin dynamics and invoking the interaction with a third atom C. In fact, we forcefully and instantly decorrelate the dynamics at the third-collision partner (atom C). However, it is expected that atom C will gradually (through further collisions) extract information from the A-B state \cite{Terashima1,Terashima2,Terashima3}, rendering $T_{\rm se}$ a lower bound for the $1/e$ entanglement lifetime.

Closing, we outline how the bipartite entanglement considered herein can manifest itself experimentally. Consider a spectroscopic measurement of the collective spin of $N$ atoms, $F_x=\sum_{j=1}^{N}f_{x}^j$, where $f_{x}^{j}$ the $x$-component of the $j$-th atom total spin. The spin noise variance, which is in principle readily measurable in spin-noise experiments, is given by $(\Delta F_x)^2=\langle F_x^2\rangle-\langle F_x\rangle^2=\sum_{j=1}^{N}(\Delta f_x^j)^2+\sum_{i\neq j}C_{f_xf_x}^{ij}$, where $C_{f_xf_x}^{ij}=\langle f_x^i f_x^j\rangle-\langle f_x^i\rangle\langle f_x^j\rangle$. Clearly, for uncorrelated atoms it is $C_{f_xf_x}^{ij}=0$, and the total spin variance equals the sum of the spin variances of the individual atoms.
Now, it is seen that a nonzero $C_{f_xf_x}^{ab}$ is connected with the entanglement produced by an SE collision between atoms A and B. Indeed, using the post-collision state $\rho$ of Eq. \eqref{rho} we can find both terms entering $C_{f_xf_x}^{ab}$. It is $\langle f_x^af_x^b\rangle=\tr\{\rho {\it f_x}\otimes {\it f_x}\}$, $\langle f_x^a\rangle=\tr\{\rho_a {\it f_x}\}$, and $\langle f_x^b\rangle=\tr\{\rho_b {\it f_x}\}$, with $\rho_a=\tr_{B}\{\rho\}$ and $\rho_b=\tr_{A}\{\rho\}$. From Fig. \ref{corr}a it is evident that a large negativity ${\cal N}(\rho)$ is connected with a large $|C_{f_xf_x}^{ab}|$. A similar result, the correlation $|C_{s_xI_x}^{ab}|$ of the electron spin of atom A with the nuclear spin of atom B, is shown in Fig. \ref{corr}b. This one is less obvious how it can be experimentally accessed, nevertheless both results clearly allude to a wealth of possible quantum correlation effects and their potential impact in precision measurements. It is thus clear that the bipartite entanglement quantified herein directly affects measurable collective spin variances.

Concluding, we have explored the atom-atom entanglement generated by spin-exchange collisions in hot alkali vapors. Our results should be equally applicable to alkali-noble gas collisions and even to cold alkali-alkali collisions, and have the potential to further advance the understanding and design of non-trivial collective quantum states for use in quantum technology. 

Both authors acknowledge useful comments by the anonymous Referees. K.M. acknowledges the co-financing of this research by Greece and the European Union (European Social Fund- ESF) through the Operational Programme «Human Resources Development, Education and Lifelong Learning» in the context of the project "Strengthening Human Resources Research Potential via Doctorate Research” (MIS-5000432), implemented by the State Scholarships Foundation (IKY)".


\begin{thebibliography}{0}
\bibitem{notation}
Regarding notation, operators like $\mathbf{s}_A\cdot\mathbf{s}_B$ imply a tensor product $\sum_{j}s_j\otimes s_j$, the left (right) part of which refers to atom A (B), and $s_j$ are the electron spin components, $j=x,y,z$. The tensor product space has dimension $2(2I_a+1)2(2I_b+1)$, where $I_a$ and $I_b$ are the nuclear spins of the colliding partners, which could be of the same or different species.
\bibitem{note0}
The time dependence of $V_1$ is implicit in its dependence on the internuclear distance, which changes with time along the collision trajectory.
\bibitem{note1}
For ease in the following, instead of $s_j$ we will use $\sigma_j=2s_j$. The particular form of Eq. \eqref{rho} is chosen to facilitate the entanglement calculations, and in any case is identically equal to existing derivations, see e.g. \cite{Happer_Primer,Happer_Book}.
\bibitem{note}
We can also create random mixed states by starting from random pure states and applying a random completely positive map, i.e. choose a set of operators satisfying $\sum_j M_j^\dagger M_j=\mathbb{1}$, and calculate $\sum_j M_j\ket{\psi}\bra{\psi}M_j^\dagger$. For example, we can use the set $M_j=\alpha_j\sigma_j$, with random $\alpha_j\geq 0$, $j=x,y,z$, and $\sum_j\alpha_j^2=1$. As expected, we obtain negativities on average smaller than for pure pre-collision states.
\end{thebibliography}
\end{document}